# First-principles based modeling of hydrogen permeation through Pd-Cu alloys

Lin Qin and Chao Jiang[*]

*State Key Laboratory of Powder Metallurgy, Central South University, Changsha, Hunan 410083, China*

**Abstract**

The solubility and diffusivity of hydrogen in disordered $Pd_{1-x}Cu_x$ alloys are investigated using a combination of first-principles calculations, a composition-dependent local cluster expansion (CDLCE) technique, and kinetic Monte Carlo simulations. We demonstrate that a linear CDCLE model can already accurately describe interstitial H in $Pd_{1-x}Cu_x$ alloys over the entire composition range ($0 \leq x \leq 1$) with accuracy comparable to that of direct first-principles calculations. Our predicted H solubility and permeability results are in reasonable agreement with experimental measurements. The proposed model is quite general and can be employed to rapidly and accurately screen a large number of alloy compositions for potential membrane applications. Extension to ternary or higher-order alloy systems should be straightforward. Our study also highlights the significant effect of local lattice relaxations on H energetics in size-mismatched disordered alloys, which has been largely overlooked in the literature.

[*]Corresponding author: email: chaopsu@gmail.com, phone: 1-806-890-4617

Present address: Department of Material Sciences and Engineering, University of Wisconsin, Madison, WI 53706, USA

## 1. Introduction

Due to their excellent $H_2$ selectivity and permeability, palladium-based membranes are widely used for the separation and purification of hydrogen, a clean fuel alternative for our future energy [1]. In its pure form, Pd membrane is susceptible to embrittlement during H loading and unloading cycles due to a α↔β phase transformation, and the associated lattice expansion and contraction [2]. Furthermore, deactivation of Pd membrane can occur after prolonged operation due to sulfur poisoning [3]. One effective route to enhance the performance of pure Pd membrane is through the addition of substitutional alloying elements, e.g., Cu, Ag, and Au [2-8]. The ultimate design goal is to identify binary or ternary alloy compositions that simultaneously exhibit high $H_2$ permeability, resistance to poisoning by gas contaminants, and structural stability against H-induced embrittlement. To this end, quantitative computational models that enable rapid and accurate prediction of hydrogen permeability in Pd-based alloys over a wide composition range are highly valuable since experimental screening of potential alloy compositions can be both costly and time-consuming.

## 2. Calculation methods

The ability of H to pass through a membrane is quantified by permeability, which can be estimated from the product of H diffusivity and solubility, assuming that bulk diffusion is the rate-limiting process and that H solubility obeys the Sievert's law [5]. While the solubility of H in a disordered alloy can be directly calculated using the supercell approach, kinetic Monte Carlo (KMC) simulations are necessarily performed to obtain H diffusivity [5-6, 8-9]. Unlike in pure metals and ordered compounds, there exist many structurally distinct binding sites and diffusion pathways in disordered alloys. For instance, in a disordered fcc $A_{1-x}B_x$ alloy, there can be $n_{NN}$ ($0 \leq n_{NN} \leq 6$) and $n_{NNN}$ ($0 \leq n_{NNN} \leq 8$) B atoms in the nearest-neighbor and next-nearest-neighbor shell surrounding an octahedral

(O) interstitial site, respectively. For tetrahedral (T) sites, we have $0 \leq n_{NN} \leq 4$ and $0 \leq n_{NNN} \leq 12$, respectively. While first-principles calculations based on density functional theory (DFT) can be performed to accurately obtain the hopping rates and transition states (TSs) of a few of those paths, exploring every path encountered in KMC simulations ("on-the-fly") is computationally prohibitive. Kamakoti and Sholl [6] first addressed this computational difficulty by developing a lattice model to correlate the binding energy of H at a given site with its local environment. In their study, H binding energy at an O (T) site was approximately described by a simple linear function of $n_{NN}$, $n_{NNN}$, and lattice parameter $a_0$. The activation barrier for O-T hop was estimated from the energy difference between T and O sites, plus some corrective terms [4, 6]. This model was subsequently refined by Sonwane et al. [7, 8], who introduced a nonlinear quadratic function to estimate H binding energies at O, T, and TS sites in Pd-Ag and Pd-Au alloys. Good agreement between theory and experiments was observed in their studies.

For bulk alloys, the dependence of their properties on lattice configuration (i.e., the spatial arrangement of atoms on an underlying parent lattice) can be characterized very efficiently by the cluster expansion (CE) technique [10-13]. Studies by Van der Ven et al. [14] and Flecha et al. [15] further showed that the CE formalism can be reliably extended to describe local environmentally-dependent properties, e.g., vacancy formation energies and H binding energies in disordered alloys. Recently, the CE method has been applied to predict H permeability in Pd-based ternary alloys [16, 17].

Motivated by the recent study of Alling et al. [18], here we further propose a concentration-dependent local cluster expansion (CDLCE) model and apply it to predict the H permeability in disordered fcc $Pd_{1-x}Cu_x$ alloys over the entire composition range. Our goal is to develop a model that combines the accuracy and robustness of the CE method [15] with the computational efficiency (i.e., applicability to arbitrary alloy compositions) of the lattice models of Kamakoti et al. [5, 6] and Sonwane et al. [7, 8]. In our CDCLE model, we characterize the local environment σ by assigning a Pd (Cu) atom around a given interstitial site a pseudo-spin variable of $\hat{S}_i = -1$ (+1). The classical H binding energy (see Ref. [6] for definition) at an O (T) site with local environment σ can then be calculated using the following Ising-like Hamiltonian:

$$E_b^{CE}(\mathbf{S}, x) = J_0(x) + J_1^{NN}(x)\sum_{i=NN}\hat{S}_i + J_1^{NNN}(x)\sum_{i=NNN}\hat{S}_i + \frac{1}{2}\sum_{i,j}J_{i,j}(x)\hat{S}_i\hat{S}_j \\ + \frac{1}{3!}\sum_{i,j,k}J_{i,j,k}(x)\hat{S}_i\hat{S}_j\hat{S}_k + \mathbf{K}$$

(1)

where $x$ is the overall alloy composition. $J(x)$ is the composition-dependent effective cluster interaction and is assumed to be a linear function of $x$ here. For the sake of simplicity, only metal atoms in the NN and NNN shells are considered. While it is possible to include more distant neighboring shells as a part of the local environment, doing so will significantly increase the number of cluster interactions that one needs to fit.

The unknown parameters in Eq. (1) are determined by least-squares fitting to DFT binding energies in $Pd_{1-x}Cu_x$ alloys with $x$ ranging from 0 to 1. The fully disordered $Pd_{0.75}Cu_{0.25}$ and $Pd_{0.5}Cu_{0.5}$ alloys are mimicked using 64-atom special quasirandom structure (SQSs) [19-23] generated using Monte Carlo simulated annealing. SQS for $x$=0.75 is obtained simply by switching the Pd and Cu atoms in the SQS for $x$=0.25. Our SQSs reproduce the pair correlation functions of perfectly random fcc alloys accurately up to the eighth-nearest neighbor at $x$=0.5 and fifth-nearest neighbor at $x$=0.25, respectively. We also ensure that the distribution of local environments in the SQSs closely follows the binomial distribution. Calculations are performed using the all-electron projector augmented wave method within the generalized gradient approximation [24], as implemented in VASP [25]. 64-atom supercells are used throughout our calculations such that the shortest distance between a H atom and its periodic images is $2\sqrt{2}a_0$. The plane-wave cutoff energy is set at 341.5 eV. For structural relaxations, we employ a 3×3×3 Monkhorst–Pack $k$-point mesh. All internal atomic positions are fully optimized using a conjugate gradient method until forces are less than 0.02 eV/Å. Upon completion of relaxation, a static calculation is performed using the linear tetrahedron method and a denser 5×5×5 $k$-point mesh to obtain accurate total energies. We find the TSs for H diffusion between neighboring O and T sites using the climbing image nudged elastic band (CI-NEB) technique [26, 27] with two intermediate images. The normal-mode vibrational frequencies of H are obtained from the eigenvalues of the Hessian matrix constructed using finite differences with a small displacement of 0.15Å. All metal atoms are rigidly constrained during such calculations.

## 3. Results and discussion

Before we go on to discuss our results, it is worthwhile to first validate our DFT calculations. In accordance with experiments [28] and previous calculations [6], we find that the lattice parameter of $Pd_{1-x}Cu_x$ decreases monotonically with $x$ and exhibits a small positive deviation from the Vegard's law. Overall, our calculated lattice constants of $Pd_{1-x}Cu_x$ alloys agree with experimental data to within 1.3%. From the total energy difference between disordered and B2-ordered PdCu, and neglecting the effect of vibrational entropy, we estimate the fcc-B2 transition temperature in Pd-Cu to be 683K, in reasonable agreement with the experimentally measured temperature of 871K [28]. The zero-point energy (ZPE) of H in the O site of pure Pd is calculated to be 0.084 eV, which agrees well with the experimental value of 0.07 eV [4].

Fig. 1 gives a direct comparison between DFT calculated and model predicted H binding energies at O and T sites in $Pd_{1-x}Cu_x$ alloys. The corresponding pair and three-body interactions are illustrated in the insets. It can be seen that CDLCE accurately reproduces DFT results over the entire composition range (0≤$x$≤1). The average fitting errors for O and T sites are 0.018 eV and 0.028 eV, respectively. While not shown, we have also used CDLCE to fit the ZPEs of H atoms, and the fitting errors are only 0.003 eV for O sites and 0.005 eV for T sites. In Fig. 1, a clear trend can be observed that the binding energies shift towards higher values as Cu content increases, which is due to the contraction of the fcc lattice by Cu addition and the resulting compression of all metal-H bonds.

Following Kamakoti et al. [5], we calculate the H solubility at a given temperature $T$ and pressure $P$ by equating the chemical potential of H in the alloy with that in the ideal $H_2$ gas, which can be written separately as follows:

$$m_H^{alloy} = E_b - \frac{1}{b}\ln\left(\frac{e^{-h(n_1+n_2+n_3)b/2}}{(1-e^{-hn_1 b})(1-e^{-hn_2 b})(1-e^{-hn_3 b})}\right) + \frac{1}{b}\ln\frac{q}{1-q} \quad (2)$$

$$m_H^{gas} = \frac{1}{2}m_{H_2}^{gas} = -\frac{1}{2b}\ln\left(\frac{1}{Pb}\left(\frac{2pm_{H_2}}{h^2 b}\right)^{3/2}\frac{e^{-hn_{H_2}b/2}}{1-e^{-hn_{H_2}b}}\frac{4p^2 I}{h^2 b}\right) \quad (3)$$

where $b = 1/k_B T$, $k_B$ is Boltzmann's constant, and $h$ is Planck's constant. $n_i$ ($i$=1-3) and $n_{H_2}$ are the vibrational frequencies of H at an interstitial site and in the gaseous state, respectively. $I = \frac{1}{4}m_{H_2}R_0^2$ is the molecular moment of inertia. $m_{H_2}$ and $R_0$ are the mass and equilibrium bond length of $H_2$ molecule, respectively. The total H solubility can then be calculated by summing $q$ values for each site in the alloy. Note that, in fcc Pd-Cu alloys, since T sites have much higher binding energies and ZPEs than O sites, their contributions to H solubility can be safely neglected. Our model predicted solubility versus temperature results are plotted in Fig. 2(a) together with experimental data [29]. The overall agreement is quite satisfactory considering that H-H interactions have been completely neglected in our calculations, which can be important for non-dilute H concentrations. For comparison, we have also performed first-principles supercell calculations [5] to obtain the H solubilities in $Pd_{1-x}Cu_x$ alloys at compositions $x$=0, 0.25, 0.5, 0.75, and 1, respectively. As shown in Fig. 2(b), CDLCE yields results in excellent agreement with direct DFT calculations over the whole composition and temperature range, which further demonstrates the predictive power of our model.

The H diffusivities in Pd-Cu alloys are obtained using KMC simulations driven by CDLCE. H diffusion is described as a series of hops between adjacent O and T sites. Note that, in order for direct O-to-O jump to occur, a H atom has to squeeze through a crowdian position between two nearest-neighbor metal atoms, which is energetically very unfavorable. We estimate the hopping rates from O to T sites using quantum corrected harmonic transition state theory [4-6]:

$$k_{O-T} = \Gamma_{O-T} \exp(-E_a / k_B T) \quad (4)$$

where $E_a$ is the classical migration barrier defined as the energy difference between TS and O sites. The prefactor $\Gamma_{O-T}$ can be calculated from the H vibrational frequencies at O and TS sites as follows:

$$\Gamma_{O-T} = \frac{\sum_{i=1}^{3} n_{O,i} f(h n_{O,i} / 2 k_B T)}{\sum_{j=1}^{2} n_{TS,j} f(h n_{TS,j} / 2 k_B T)} \quad (5)$$

where $f(x) = \sinh(x)/x$. At a given temperature, the numerator and denominator in Eq. (5) are fitted separately using CDLCE. The hopping rate from T to O can be calculated similarly. We employ a 20×20×20 cubic simulation cell (with periodic boundary conditions) containing 32000 sites, which are randomly occupied by Pd and Cu atoms for a desired overall alloy composition. We then randomly insert 300 H atoms and equilibrate the system for 100000 MC steps. An additional 1000000 MC steps are performed and the H diffusivity is obtained from the long-time limit of the mean-square displacement according to the Einstein relation. Our final results are summarized in Fig. 3. Similar to solubility, the H diffusivity in Pd-Cu also decreases with increasing Cu content, although such a decrease becomes less dramatic with increasing temperature. Such a conclusion is fully consistent with previous studies [5, 6]. Our calculated permeability results are also in reasonable agreement with experiments [30, 31].

At low temperatures, Pd and Cu atoms on the fcc lattice may undergo long-range ordering (LRO) to form superstructures. Even without the emergence of LRO, short range order can still persist [12, 13]. Since our CDLCE model is parameterized entirely using DFT calculations on fully disordered alloys, its applicability to ordered systems remains in question. As a test case, we consider the O site in the $L1_2$ ordered $Pd_3Cu$ compound that is surrounded by six NN Pd atoms and eight NNN Cu atoms. As shown in Fig. 1(a), the model predicted H binding energy with such a local environment is ~0.14 eV lower than that from DFT, although the overall lattice parameters of ordered and disordered $Pd_3Cu$ are almost identical. To understand the origin of such an apparent discrepancy, we obtain the distributions of O-site volumes in random fcc Pd-Cu alloys from the fully relaxed 64-atom SQSs. As shown in Fig. 4, there exist many distinct O-site volumes, all deviating strongly from that of the average lattice, i.e., $\bar{V}_O = \frac{1}{6}a_0^3$. This is a clear indication of significant local lattice relaxations (i.e., local displacements of atoms from their ideal lattice positions) in disordered Pd-Cu alloys, which has also been previously reported by Lu et al. [32]. With increasing Cu content, all O-site volumes decrease at approximately the same rate. At a given alloy composition, the O-site volume monotonically increases with increasing number of Pd atoms in the NN shell. Interestingly, for O sites with 6 NN Pd atoms, their volumes are actually closer to the equilibrium O-site volume in pure Pd than to the lattice average, which favors H binding. In contrast, in $L1_2$-ordered $Pd_3Cu$, all O-site volumes are equal to $\bar{V}_O$ since local lattice relaxations are now completely forbidden by symmetry. Clearly, even for the same alloy composition, O sites with similar local environments can have very different volumes (and thus different H binding energies) depending on the state of order of the alloy. Presumably, this can explain the large discrepancy mentioned above.

## 4. Concluding remarks

In this study, we propose a composition dependent local cluster expansion model to describe the dissolution and diffusion of interstitial H in disordered Pd-Cu alloys over the entire range of compositions. The accuracy of our model is validated by comparing predicted results with experimental measurements as well as DFT calculations. Remarkably, in describing the composition dependence of the effective cluster interactions in Eq. (1), even a simple linear function can adequately capture most of the important physics. Indeed, our tests suggest that introducing non-linearity in $J(x)$ does not lead to significant improvement in accuracy, although the number of fitting parameters is increased considerably. We also find that local lattice relaxations have a profound effect on hydrogen energetics in size-mismatched disordered alloys. Such relaxations are greatest in extent in random alloys, but can be completely forbidden due to symmetry restrictions in ordered compounds. We therefore speculate that it may be challenging to model H in the disordered and ordered states of an alloy on equal footing using the "local environment" concept.

## Acknowledgements

This work is financially supported by the National Natural Science Foundation of China (Grant No. 50901091). Support from the "New Century Outstanding Talent Supporting Program" of Chinese Ministry of Education is also acknowledged.

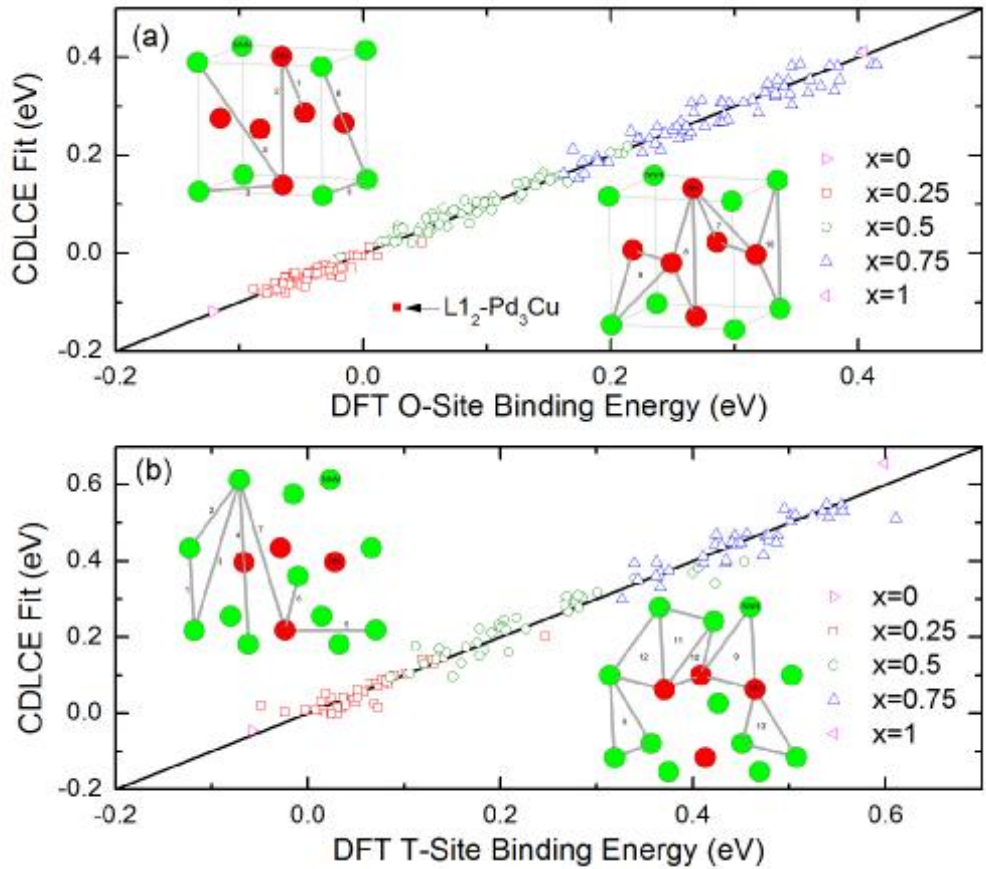

FIG. 1. Comparisons between DFT calculated and CDLCE predicted classical H binding energies at octahedral (a) and tetrahedral (b) interstitial sites in fully disordered fcc $Pd_{1-x}Cu_x$ alloys at compositions $x$=0, 0.25, 0.5, 0.75, and 1, respectively. The solid line represents perfect agreement between those two methods.

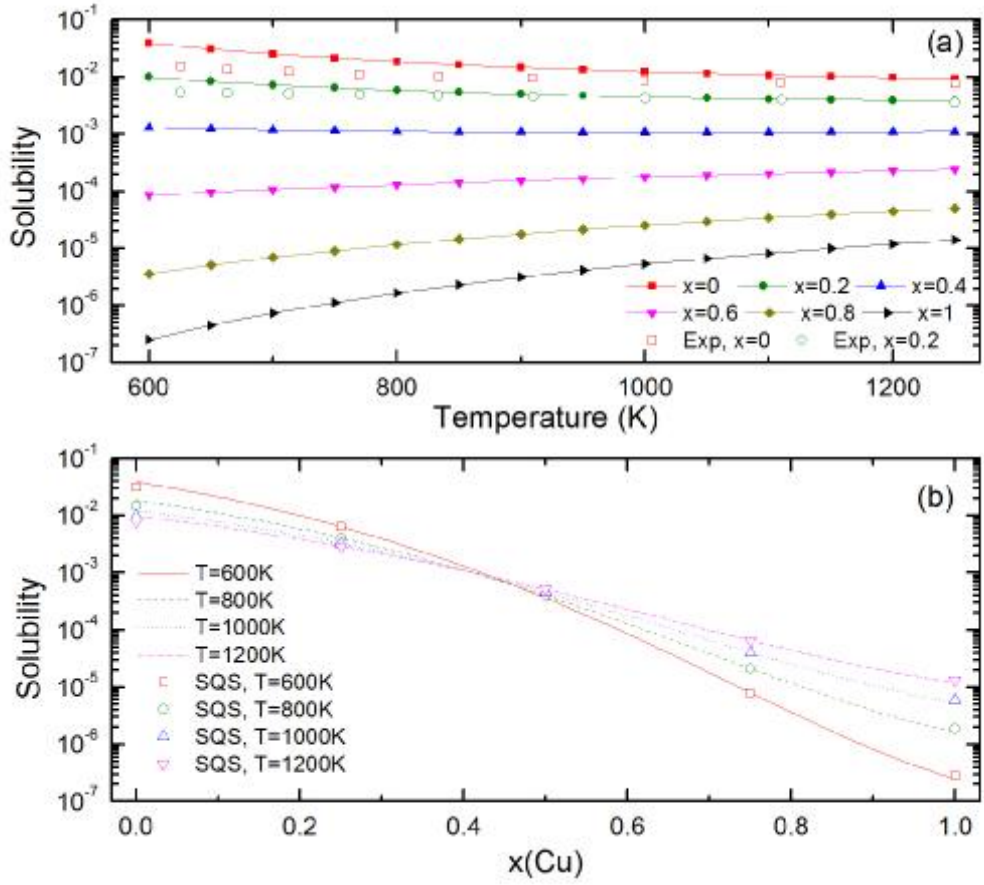

FIG. 2. CDCLE predicted H solubility (solid, dashed, and dotted lines) in random $Pd_{1-x}Cu_x$ at 1 atm in comparison with experimental measurements [29] (a) and direct DFT calculations (b).

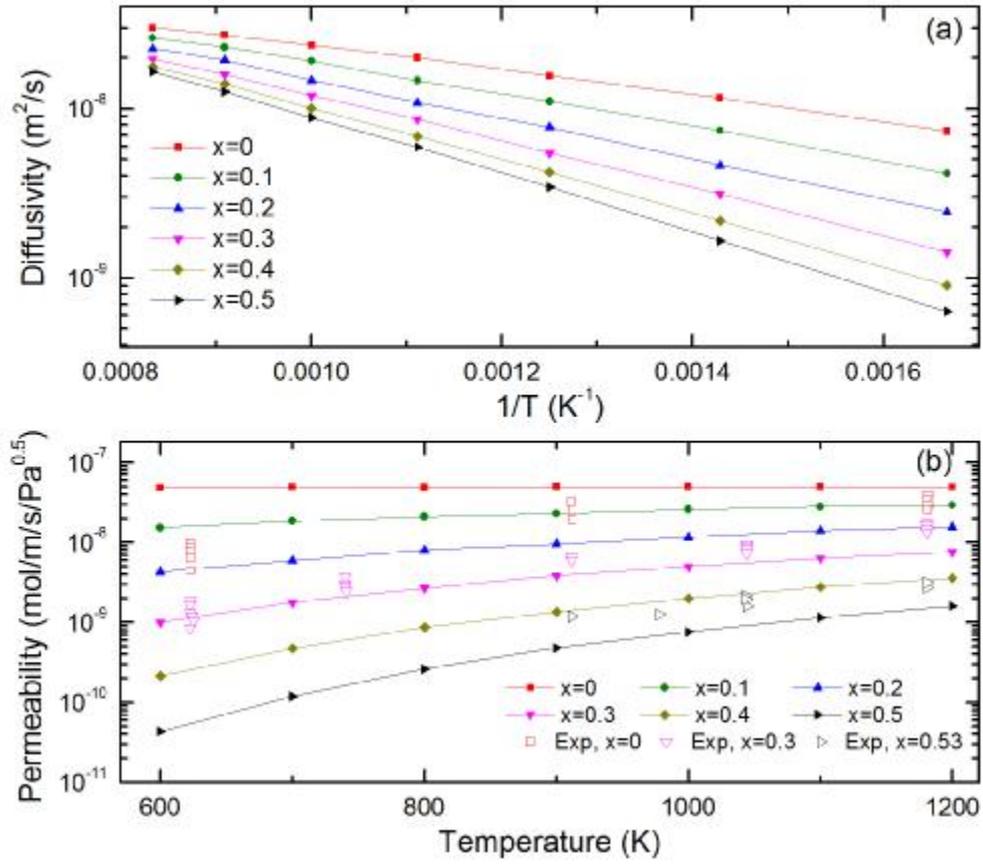

FIG. 3. Model predicted diffusivity (a) and permeability (b) of H in $Pd_{1-x}Cu_x$ (full symbols). Experimental permeability data [30, 31] are shown as open symbols.

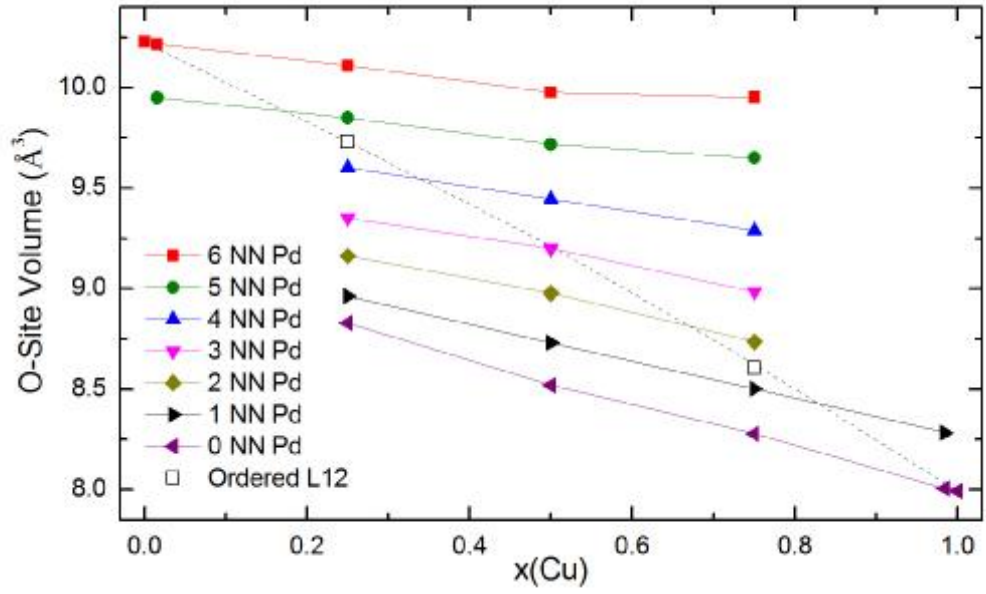

FIG. 4. SQS calculated average O-site volumes in random fcc $Pd_{1-x}Cu_x$ alloys, categorized according to the number of Pd atoms in the NN shell. The dashed line represents the octahedron volume of the average lattice. The O-site volumes in $L1_2$-ordered compounds are all identical by symmetry (open symbols).